\documentclass[aps,prl,twocolumn,showpacs]{revtex4}
\usepackage{bm}
\usepackage{graphicx}
\usepackage{amsmath}
\usepackage{eufrak}
\usepackage{color}
\newcommand{\nix}[1]{}
\begin{document}

\title{Spin-polarized electric currents in diluted magnetic 
semiconductor heterostructures\\
 induced  by terahertz and microwave radiation}
\author{P. Olbrich,$^1$
C. Zoth,$^1$ P. Lutz,$^1$ C. Drexler,$^1$ V. V. Bel'kov,$^2$ Ya.
V. Terent'ev,$^2$ S. A. Tarasenko,$^{2}$ A. N. Semenov,$^{2}$ S.
V. Ivanov,$^{2}$ D. R. Yakovlev,$^{2,3}$
T. Wojtowicz,$^4$ U. Wurstbauer,$^5$ D. Schuh,$^1$ and S.
D.\,Ganichev$^{1}$}
\affiliation{$^1$  Terahertz Center, University of Regensburg, 93040
Regensburg, Germany}
\affiliation{$^2$Ioffe Physical-Technical Institute, Russian
Academy of Sciences, 194021 St.\,Petersburg, Russia}
\affiliation{$^3$ Experimental Physics 2, TU Dortmund University,
44221 Dortmund, Germany,}
\affiliation{$^4$ Institute of  Physics, Polish Academy of
Sciences, 02668 Warsaw, Poland,}
\affiliation{$^5$ Columbia University, 435 W 116th St New York, NY
10027-7201, USA}

\begin{abstract}
We report on the study of spin-polarized electric  currents in
diluted magnetic semiconductor (DMS) quantum wells subjected to an
in-plane external magnetic field and illuminated by microwave or
terahertz  radiation. The effect is studied in (Cd,Mn)Te/(Cd,Mg)Te
quantum wells (QWs) and (In,Ga)As/InAlAs:Mn QWs belonging to the
well known II-VI and III-V DMS material systems, as well as, in
heterovalent AlSb/InAs/(Zn,Mn)Te QWs which represent a promising
combination of II-VI and III-V semiconductors. Experimental data
and developed theory demonstrate that the photocurrent originates
from a spin-dependent scattering of free carriers by static
defects or phonons in the Drude absorption of radiation and
subsequent relaxation of carriers. We show that in DMS structures
the efficiency of the current generation is drastically enhanced
compared to non-magnetic semiconductors. The enhancement is caused
by the exchange interaction of carrier spins with localized spins
of magnetic ions resulting, on the one hand, in the giant Zeeman
spin-splitting, and, on the other hand, in the spin-dependent
carrier scattering by localized Mn$^{2+}$  ions
polarized by an external magnetic field.
\end{abstract}
% 73.50.Pz Photoconduction and photovoltaic effects
% 72.25.Fe Optical creation of spin polarized carriers
% 72.25.Rb Spin relaxation and scattering
% 78.67.De Quantum wells
\maketitle

\section{I. Introduction}

Transport of spin-polarized carriers in low-dimensional
semiconductor structures is in the focus of intensive research
aiming at 
spintronics~\cite{Maekawa,Fabian,spinbook,spintronics2008,Awschalombook2010,Wu2010,Tsymbal}. 
In particular, spin transport phenomena in diluted magnetic
semiconductors (DMS) are currently discussed as a key issue for the development of 
semiconductor based spintronic devices, see e.g. Ref.~[\onlinecite{Awschalombook2010,DMS2010,handbook2010,Ohno2010,Dietl2010,FlatteNature2011}].
DMS materials represent semiconductors where paramagnetic ions, usually Mn, 
are introduced in the host III-V or II-VI materials~\cite{Fur88}. 
The magnetic properties of
the DMS structures can be widely tuned from
paramagnetic to ferromagnetic behavior by varying concentration of magnetic ions, their location in 
the heterostructure and by the structure fabrication,
Strong ''sp-d'' exchange interaction, which couples free carrier spins with the localized spins of magnetic ions, greatly enhances magneto-optical 
and magneto-transport effects in DMS structures.
An important issue in the field of spin-dependent phenomena 
is the generation of spin currents 
or spin polarized electric currents, e.g. due to electric 
spin injection, anomalous Hall effect, spin Hall effect,  and spin polarized tunneling.
A further way to generate spin-polarized currents provides a spin dependent scattering 
of free carriers excited by infrared or terahertz (THz) radiation. 
This effect was observed in various low dimensional non-magnetic semiconductor structures~\cite{handbook2010,Nature06,PRBSiGe,Shen10}
and has been shown to be strongly enhanced in DMS structures~\cite{PRL09}
The advantage of DMS structures a nearly fully spin polarized electric current may be generated 
due to the strong ''sp-d'' exchange interaction.

In this paper we give a detailed theoretical description  of spin
current mechanisms  in DMS heterostructures. Experimental results
are presented for DMS structures based on II-VI and III-V
semiconductors as well as for hybrid II-VI/III-V heterostructures.
We shown that the exchange interaction in DMS structures yields
two roots of spin-polarized current generation. The one is related
to the direct effect of a magnetic field on free carriers (the
Zeeman splitting with intrinsic $g$-factor) and another one to the
strong exchange interaction (scattering) of free carriers with
magnetic Mn$^{2+}$ ions.

The paper is organized as follows. In Sec.~II, we present a
microscopic theory of optically-induced spin-polarized currents in
DMS structures. We discuss corresponding models and the current
behavior upon the variation of parameters of optical excitation
(photon energy and polarization), sample characteristics and
temperature. Section~III describes the experimental technique and
geometry of measurements as well as radiation sources used. In
Sec.~IV we  describe the design and parameters of the samples,
present experimental data and compare the results with theory. We
start with the well known DMS material QW systems based on
$n$-(Cd,Mn)Te/(Cd,Mg)Te (Sec.~IV~A) and $p$-(In,Ga)As/InAlAs:Mn (Sec.~IV~B) and then
introduce the results obtained on recently designed heterovalent
hybrid AlSb/InAs/(Zn,Mn)Te structures with a two-dimensional
electron gas. The paper is summarized in Sec.~V.

\section{II. Microscopic  model}

\begin{figure}
\includegraphics[width=0.95\linewidth]{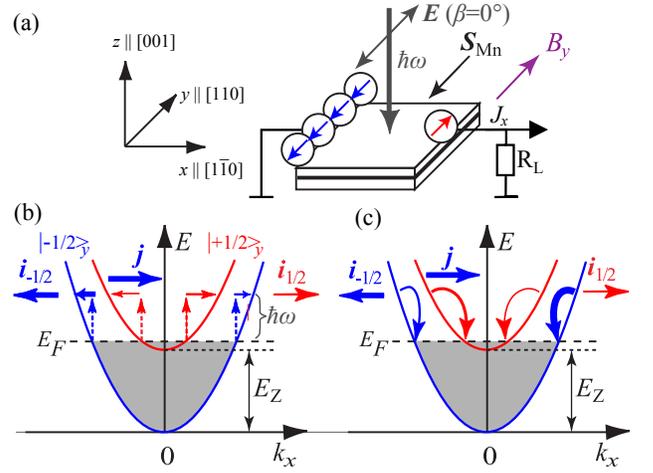}
\caption{  Model of spin-polarized electric
currents induced by terahertz/microwave radiation in  DMS  
QW structures subjected to an in-plane
magnetic field. (a) Illustration of the transverse electric
current induced by the linearly polarized radiation at normal
incidence and caused by the Zeeman  splitting.
The figure also sketches the typical experimental geometry
 where the
electric current is measured by the voltage drop over
 the load resistance, $R_L$.
 Arrows show directions
of radiation electric field vector, $\bm E$, magnetic field $B_y$
and average spin of Mn$^{2+}$ ions, $\bm S_{Mn}$. Circles
with oppositely directed arrows 
 show electrons with opposite spins. 
(b) and (c)  Excitation and relaxation mechanisms of
the current generation. Due to spin-dependent scattering the
transitions to the states with positive and negative $k_x$ in the
spin subbands occur at different rates, which leads to the
oppositely directed electron fluxes $\bm{i}_{1/2}$ and
$\bm{i}_{1/2}$. This is illustrated for (b) scattering-assisted
Drude absorption and (c) energy relaxation processes. The Zeeman
splitting of the spin subbands results in their non-equal
population. It disturbs the balance between the fluxes
$\bm{i}_{1/2}$ and $\bm{i}_{1/2}$ giving rise to net spin
polarized electric current. 
} \label{Fig01}
\end{figure}

The origin of spin-polarized current generation is
spin-dependent scattering
of free carriers by static defects or phonons at the Drude
absorption of radiation  and subsequent relaxation of
carriers~\cite{handbook2010}. This is due to
spin-orbit interaction in gyrotropic media,
such as InAs-, GaAs-, and CdTe-based two-dimensional structures,
 that gives rise to linear in the wave vector terms
in the matrix element of scattering. The total matrix element of
scattering can be thus presented by~\cite{Ivchenko08SST}
\begin{equation}\label{spin_scattering}
V_{\bm{k}'\bm{k}} = V_0 + \sum_{\alpha\beta} V_{\alpha\beta}
\sigma_{\alpha} (k_{\beta} + k^\prime_{\beta}) \:,
\end{equation}
where the first term on the right-hand side describes the
conventional spin-independent scattering, $\sigma_{\alpha}$ are
the Pauli matrices, $\bm{k}$ and $\bm{k^\prime}$ are the initial
and scattered wave vectors, and $\alpha$ and $\beta$ are the
Cartesian coordinates.
The linear in the wave vector contributions stem from bulk and
structure inversion asymmetry of QWs. The spin and electron
momentum dependent scattering results in an asymmetry of electron
distribution in $\bm{k}$-space in each spin
subbands if the electron gas is driven out of equilibrium. The
corresponding processes for the Drude absorption, which is
accompanied by scattering, and energy relaxation are illustrated
in Figs.~\ref{Fig01}(b) and~\ref{Fig01}(c), respectively. Thus,
 the spin-dependent scattering leads to the emergence
of oppositely directed electron fluxes $\bm{i}_{\pm1/2}$ in the
spin subbands.
For zero magnetic field the fluxes are of equal magnitude forming a
pure spin current defined as
$\bm{J}_{s} = (1/2)  (\bm{i}_{+1/2} -
\bm{i}_{-1/2})$. At nonzero magnetic
field  $\bm{B}$, the fluxes of electrons with the
spin  projections $\pm 1/2$ along $\bm{B}$ become unbalanced
giving rise to a net electric current  $\bm{j} = e
(\bm{i}_{+1/2} + \bm{i}_{-1/2})$, where $e$ is the electron
charge,
see Figs.~\ref{Fig01}(a)-(c).  The microscopic calculation of the
fluxes $\bm{i}_{\pm1/2}$ based on the Boltzmann approach is given in Appendix.

A straightforward mechanism causing the electric current is 
the unequal population of the
spin subbands due to the Zeeman effect.
The mechanism
is sketched in Figs.~\ref{Fig01}(b) and~\ref{Fig01}(c)  for the
scattering-assisted optical excitation (\textit{excitation mechanism}) and
relaxation (\textit{relaxation mechanism}), respectively. In the case of
photoexcitation, Fig.~\ref{Fig01}(b), the transition rate in each
spin subband depends on the subband population $n_{\pm 1/2}$.
Consequently, the electron fluxes in the spin subbands
$\bm{i}_{\pm 1/2} \propto n_{\pm 1/2}$ become unequal resulting in
the electric current
\begin{equation}\label{Eq01}
\bm{j}_{\rm Z} = 4 e s(B) \bm{J}_{s} \:.
\end{equation}
where $s(B)$ is the average electron spin  projection along $\bm{B}$.
For a low degree of spin polarization, it is given by
\begin{equation}\label{Eq02}
s(B) = \frac12 \frac{n_{+1/2}-n_{-1/2}}{n_{+1/2}+n_{-1/2}} =  -  \frac{E_{\rm Z}}{4 \bar{E}} \,.
\end{equation}
Here, $E_{\rm Z}$ is the Zeeman splitting energy and $\bar{E}$ is
the characteristic electron energy, equal to  the
Fermi energy $E_{\rm F}$ for degenerate 
two-dimensional  electron gas (2DEG) and $k_{\rm B} T$
for non-degenerate gas  at the temperature $T$, respectively. 
For linearly polarized radiation, photoexcited carriers are
preferably aligned along the radiation electric field. Therefore,
for a fixed magnetic field direction, e.g., $\bm{B}\| y$, the
polarization plane rotation, described by the azimuth angle,
 results in oscillations of the $x$ and $y$
current components as a function of $\beta$. Similarly, the Zeeman
splitting gives rise to an electric current in the case of energy
relaxation of hot electrons. Of course, the latter mechanism,
which is based on electron gas heating, is independent of the
radiation polarization. Microscopic and  symmetry
analysis show that it results in a current along the 
$x$-direction for $\bm{B}\| y$.
For (001)-oriented QWs, the polarization dependences of  the total 
transverse, $j_x$, and longitudinal, $j_y$, photocurrents  are 
given by (see Appendix)
\begin{equation}\label{Eq03}
j_{x} = j_1 + j_2 \cos 2\beta \quad \quad j_{y} = j_3 \sin 2\beta \:,
\end{equation}
where $j_1$, $j_2$ and $j_3$ are polarization independent
parameters, describing the  relaxation
mechanism ($j_1$) and excitation mechanism ($j_2$ and $j_3$),
$\beta$ is the azimuth angle ($\beta = 0$ for $\bm E || \bm B$),
and $x \| [1\bar{1}0]$ and $y \| [110]$ are the Cartesian
coordinates.  Note that while the direction of the
polarization-independent current is determined by the magnetic
field direction, QW crystallographic orientation and design, the
directions of the polarization-dependent components can be varied
just by rotation of the polarization plane. 

In diluted magnetic semiconductors, the considered mechanism of
the photocurrent generation  is drastically enhanced due to the
giant Zeeman splitting. In $n$-type DMS structures, the splitting
is given by the sum of intrinsic and exchange
contributions\,\cite{Fur88}
\begin{equation}
\label{Eq04}
E_{\rm Z} = g_{e(h)} \mu_{\rm B} B +  \bar{x} S_0 N_0 \alpha_{e(h)} {\rm B}_{5/2}
\left( {\frac{5 \mu_{\rm B} g_{Mn} B}{2 k_{\rm B} (T_{Mn} + T_0 )}} \right) \,\, ,
\end{equation}
where $g_{e(h)}$ is electron (hole) Lande factor in the absence 
of magnetic impurities, $\mu_{\rm B}$ is  the Bohr mag\-neton, $\bar{x}$ is
the effective average concentration of Mn,  $N_0 \alpha_{e(h)}$ is the
exchange integral for conduction (valence) band carriers, $g_{Mn} = 2$ is
Mn $g$-factor, $T_{Mn}$ is the Mn-spin system
temperature. Parameters $S_0$ and $T_0$ account for the Mn-Mn
antiferromagnetic interaction, and $\rm{B}_{5/2}\left( \xi
\right)$ is the modified Brillouin function.

\begin{figure}[htb!]  %15
\includegraphics[width=0.99\linewidth]{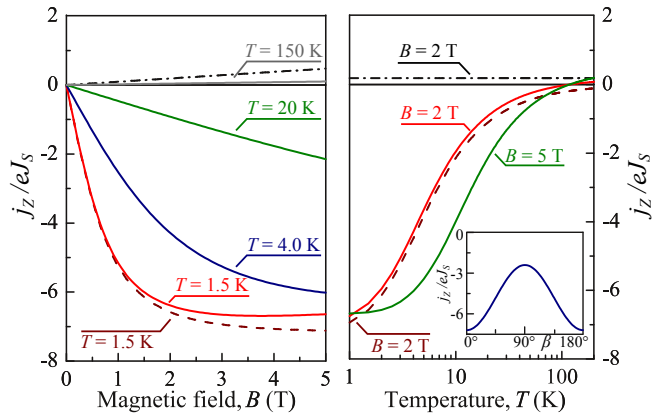}
\caption{(a) Magnetic field dependence of the photocurrent 
contributions
caused by the intrinsic and exchange Zeeman splitting and
calculated after Eqs.~(\ref{Eq01}) and~(\ref{Eq04}) for different
temperatures. (b) Temperature dependence of the photocurrent
contributions calculated for different magnetic field. Dashed 
lines correspond to the effect caused by exchange mechanism, 
dot-dashed lines reflect photocurrents driven by intrinsic one.
Curves are obtained for literature values of parameters for \textit{n}-(Cd,Mn)Te;
$g_e = - 1.64$, $\bar{x}= 0.013$, 
$N_0 \alpha_e = 220$~meV, $E_{\rm F} = 10$~meV, and $T_{Mn} = 0$, 
see Ref. [\protect \onlinecite{Fur88}]. 
Inset shows an example of 
polarization dependence of the photocurrent  given by
Eq.~(\ref{Eq03}). 
} \label{Fig02}
\end{figure}

Shown in Fig.~\ref{Fig02} are  magnetic field and temperature
dependences of the photocurrent  $j_e$ calculated
after Eqs.~(\ref{Eq01}), (\ref{Eq02}) and (\ref{Eq04}) taking into
account literature values of parameters for \textit{n}-(Cd,Mn)Te
QWs. In order to focus on the effect of magnetic impurities we
normalized the current  $j_e$ by the
pure spin current $J_s$, which depends on particular scattering
mechanism and, therefore, may depend on temperature. At low
temperatures, the current is dominated by the exchange interaction
between free electrons and magnetic ions following the Brillouin
function. As a result, the current first linearly grows with the
magnetic field strength and then saturates. The increase in temperature
leads to the decrease of the current magnitude and shifts  the
 saturation to higher fields. Finally, at high
temperatures,  the exchange contribution to the
Zeeman splitting
becomes comparable or even smaller than the intrinsic one.
In CdTe, where the sign of the intrinsic $g_e$-factor is opposite to the 
exchange one~\cite{Wojtowicz99}.
This interplay results in a change of the photocurrent sign, see Fig.~\ref{Fig02}.
In some other materials, e.g, $p$-type (In,Ga)As:Mn DMS structures,
both contributions have the same sign~\cite{spinbook} and inversion does not
occur. 

Equation~(\ref{Eq02}) yielding the linear relation between the
average electron spin  and the Zeeman splitting is valid for a low
degree of electron gas polarization only. This 
regime is relevant for the majority of structures 
at moderate magnetic fields of several Tesla. However, in DMS
structures, where high degree of spin polarization can be achieved
at moderate magnetic fields,  
the linear relation can be violated. This is another reason for the current
saturation with rising the magnetic field. Indeed, in
a fully spin-polarized electron gas, which can occur at low
temperatures in DMS for magnetic fields even well below saturation
of magnetization,
the electron flux in one of the spin subbands vanishes. Therefore,
the electric current becomes independent  of the
Zeeman splitting   and is given by $\bm{j}=\mp 2e
\bm{J}_{s}$, where the sign is determined by the effective
$g^*$-factor sign.

The described variation of the photocurrent with temperature and
magnetic field is  relevant for both the excitation and relaxation
mechanisms 
sketched in Fig.~\ref{Fig01}. However, an effective way to
distinguish between these microscopically different 
mechanisms is to study the polarization dependence
of the photocurrent. Indeed, for a fixed magnetic field, the
excitation related photocurrent varies upon rotation the radiation
polarization plane, see Eqs.~(\ref{Eq03}), while the relaxation
related current does not. An example of the dependence of the
current containing both contributions on the azimuth angle $\beta$
is shown in the inset in Fig.~\ref{Fig02}(b).

So far, we have considered  mechanisms of the spin
polarized current
formation based on the Zeeman splitting of electron spin subbands.
However, there is an additional contribution to the photocurrent
generation being specific for DMS structures. It is related to the
spin-dependent electron scattering by polarized Mn$^{2+}$ ions,
which is described by the
Hamiltonian  of interaction between band electrons 
and magnetic ions~\cite{Fur88}
\begin{equation}
\label{Eq05}
H_{e-Mn}=\sum_i \left[ u - \alpha \,
( \hat{\bm{S}}_i \cdot  \hat{\bm{s}} )
\right]
\delta(\bm{r} - \bm{R}_i) \:.
\end{equation}
Here the index $i$ enumerates Mn ions,  $\hat{\bm{S}}_i$ 
is the ion spin operator,
$\hat{\bm{s}} = \bm{\sigma}/2$ 
is the electron spin operator,
$u \delta(\bm{r} - \bm{R}_i)$ is the scattering potential 
without exchange interaction, $\bm{r}$ the electron coordinate, 
and $\bm{R}_i$ the ion position.
Note that the parameter $\alpha$ in Eq.\,(\ref{Eq05}) is also 
responsible for the giant Zeeman splitting in Eq.\,(\ref{Eq04}).

The external magnetic field polarizes the Mn spins leading to 
different scattering rates for
band electrons with the spin projection $\pm 1/2$ along the 
ion polarization~\cite{Egues}.
Accordingly, the momentum relaxation times in the spin 
subbands $\tau_{p,+1/2}$ and $\tau_{p,-1/2}$ become unequal.
Since the electron fluxes $\bm{i}_{\pm1/2}$ depend on the 
momentum relaxation times in  the
spin subbands, they do not compensate one another  giving rise
to a net electric current $\bm{j}_{Sc}$.
This photocurrent can occur even for equally populated  spin subbands
and, therefore, is superimposed on the Zeeman splitting 
related contribution $\bm{j}_Z$.
An estimation for $\bm{j}_{Sc}$ can be made
assuming that the momentum relaxation of electrons is  determined
by their interaction with Mn$^{2+}$ ions.
 Taking into account the fact
that the spin independent part of the Mn potential, characterized
by $u$, is usually much larger than the exchange term described by
$\alpha$, we obtain
\begin{equation}
\label{Eq06}
\bm{j}_{Sc} = 2 e \tau_p \frac{\alpha}{u} 
\frac{\partial \bm{J}_{s}}{\partial \tau_p} \,  S_{Mn}  \:,
\end{equation}
where $\tau_p$ is the electron momentum relaxation time for the 
case of non-polarized ions, $S_{Mn}$ is the average Mn spin  
projection along $\bm{B}$,  $S_{Mn}=-S_0 B_{5/2}(\xi)$, and $\bm{J}_{s}$ 
is formally considered as a function of $\tau_p$.
Similarly to the current caused by the giant Zeeman splitting, 
the scattering related current (\ref{Eq06}) is determined by the Mn ions 
polarization and, therefore, is characterized by non-linear magnetic 
field dependence vanishing at high temperatures.

For a low degree of electron gas polarization, the
photocurrent  is
given by the sum of two contributions
\begin{equation}
\label{Eq07}
\bm{j} = \bm{j}_{\rm Z} + \bm{j}_{Sc} \:.
\end{equation}
Due to the fact that both  terms
are caused by the exchange interaction, the resulting  electric
current will follow the Brillouin function  no matter which contribution dominates.
Depending on the structure material the currents
$\bm{j}_{\rm Z}$ and $\bm{j}_{Sc}$ may interfere in constructive
or distractive ways~\cite{footnote0}.
Possible ways to distinguish the relative
contributions of $\bm{j}_{\rm Z}$ and $\bm{j}_{Sc}$ to the total
spin-polarized electric current are to compare the temperature
behavior of the current with that of the Zeeman splitting or to
study the dependence of the current on the radiation frequency and
structure mobility.

Finally we note, that at high temperatures, where the exchange
enhancement of the current is absent, additional orbital
mechanisms may contribute to the magnetic field induced
photocurrent.
The orbital contribution comes from an asymmetry in the electron scattering 
due to the Lorentz force
acting upon carriers~\cite{Tarasenko08,Tarasenko11}.
Its sign depends on the QW design and scattering mechanism.
Therefore, the interplay of spin and orbital mechanisms may 
influence the current behavior,
 e.g., results in shifting the temperature inversion point or even 
 its appearance/disappearance.
The orbital  contribution to the photocurrent may
also show up at low  temperatures and high magnetic fields. Being
linear in the magnetic field,  it
may lead to a deviation  of the field behavior of the
measured current from the Brillouin  function expected for the
exchange mechanism.

\section{III. Experimental technique}

The experiments have been carried out on three
different types of DMS low-dimensional structures with Mn$^{2+}$
as the magnetic impurity. Here, spin-polarized photocurrents have
been studied in the well known II\,-–\,VI and III\,-–\,V DMS
 systems, represented by $n$-type (Cd,Mn)Te/(Cd,Mg)Te QWs
and $p$-type (In,Ga)As/InAlAs:Mn QWs, respectively, as well as in
heterovalent hybrid II\,-–\,VI/III\,-–\,V $n$-type
AlSb/InAs/(Zn,Mn)Te QWs  with Mn layers inserted into the
II\,-–\,VI barriers. All structures have been grown by
molecular-beam epitaxy on semi-insulating (001)-oriented GaAs
substrates with buffer layers corresponding to each material group
in order to relax strain.

A set of  $5 \times 5$~mm$^2$ sized samples from the quantum well
structures having various density and spatial position of
Mn-doping layers have been prepared. To measure the photocurrent
two pairs of ohmic contacts at the center of the sample edges
oriented along the $x
\parallel [1\bar {1}0]$ and $y
\parallel [110]$ directions have been prepared [see inset in
Fig.~\ref{Fig04}(b)]. The specific structures design and
parameters are given in the beginning of the corresponding
sections presenting the experimental results (see Sec.~IV~A-C).
The samples were placed into an optical cryostat with $z$-cut
crystal quartz windows and split-coil superconducting magnet. The
magnetic field
\textbf{\emph{B}} up to 7~T was applied in  the QWs plane along $y
\parallel [110]$ axis. The sample temperature was varied from 1.8
up to 200~K.

The experimental geometry is sketched in Fig.~\ref{Fig04}(b). The
measurements of magnetic-field-induced photocurrents are carried
out under excitation of the (001)-grown QW samples with
linearly polarized terahertz and microwave
radiation at normal incidence. The experimental arrangement is chosen to 
exclude any effects known to cause photocurrents at zero magnetic field~\cite{Ganichev02}.
For optical excitation we use four different types of radiation
sources: low power $cw$ optically pumped CH$_3$OH THz laser,  Gunn
diodes, backwards wave oscillator and high power pulsed optically
pumped THz laser. The sources provided monochromatic radiation in
the frequency range between 0.1 and $\approx$2.5~THz
(corresponding photon energies, $\hbar \omega$ varied from 0.3 up
to 10~meV). The radiation photon energies are smaller than the
band gap as well as the size-quantized subband separation. Thus,
the radiation induces indirect optical transitions in the lowest
conduction subband (Drude-like free-carrier absorption).

Low power excitation with $P \approx 2$~mW at the sample spot is
obtained by the CH$_3$OH THz laser emitting radiation with
frequency $f = 2.54$~THz (wavelength $\lambda = 118~\mu$m) ~\cite{Kvon08},
backwards wave oscillator (Carcinotron) operating at $f = 290$~GHz
($\lambda = 1.03$~mm) and a Gunn diode with $f = 95.5$~GHz
($\lambda = 3.15$~mm). The incident power of the $cw$ sources was
modulated between 255 and 800~Hz by a pin switch (Gunn diode) or
an optical chopper. The photocurrent is measured across a 1
M$\Omega$ load resistor applying the standard lock-in technique.
Pulsed high power THz radiation with  $f \approx 2.03$~THz
($\lambda = 148~\mu$m), a peak power $P$~$\approx$~40~kW at the
sample spot, and a pulse duration of $\approx$~200~ns is obtained
by a NH$_3$ laser optically pumped with a TEA CO$_2$ laser~\cite{physicaB99,Schneider2004}. 
In this set-up the signal is detected via a voltage drop over a 50
$\Omega$ load resistor applying a fast amplifier and a storage
oscilloscope. The radiation power has been controlled by either
pyroelectric detectors or THz photon drag detector.
The radiation is focused onto samples by one or
two parabolic mirrors (for  lasers and Carcinotron, respectively)
or horn antenna (Gunn diode).
Typical laser spot diameters varied, depending on the wavelength, from 1 to 3~mm. 
The spatial beam distribution had an almost Gaussian profile, 
checked with a pyroelectric camera.
The spatial distribution of the microwave radiation at 
the sample's position, and, in particular, the efficiency of the 
radiation coupling to the sample, by, e.g., 
the bonding wires and metalization of contact pads,
could not be measured. Thus, all microwaves data 
are given in arbitrary units. In order to
vary the angle $\beta$ between the light polarization plane and
the magnetic field, the plane of polarization of the radiation
incident on the sample was rotated by means of $\lambda/2$-plates.
Hereafter, the angle $\beta$ = 0$^\circ$ is chosen in such a way
that the incident light polarization is directed along the
$y$-axis, see inset in Fig.~\ref{Fig04}(b).

\section{IV. Photocurrent experiments}

In the following Sections (A-C) we present the experimental
results for three different groups of DMS low dimensional
structures. The sections are organized in a similar way; we start
with the description of the structures design/parameters, than
present a detailed study of the photocurrent behavior upon
variation of  the magnetic field strength, temperature, radiation
intensity and polarization, and, finally give a comparison of the
results with the theory described in Sec.~II.

\subsection{A. $n$-(Cd,Mn)Te/(Cd,Mg)Te quantum wells}

Low dimensional structures based on wide band gap II-VI diluted
magnetic semiconductors are the best understood DMS materials with
the most studied electric and magnetic properties~\cite{DMS2010}, and it is the DMS system
in which the terahertz radiation induced
spin-polarized electric current has been reported ~\cite{PRL09,Drexler10}. 
The experiments presented below have been
carried out on 10~nm wide $n$-type (Cd,Mn)Te single QWs embedded
in (Cd,Mg)Te barriers. The DMS QWs were grown by molecular-beam
epitaxy on (001)-oriented GaAs substrates~\cite{Egues,Crooker,Jaroszynski2002}.
Three groups of $n$-(Cd,Mn)Te/Cd$_{0.76}$Mg$_{0.24}$Te structures with different
Mn contents (A0, A1 and A2)  were fabricated.
In each group
several samples from the same wafer were investigated.
In the following we discuss the data obtained on the non-magnetic
reference QWs (samples~A0) and DMS QWs (with A0, A1 and A2
samples, discussed later being, the representative of each of
those groups) having different magnetic properties.
In samples A1 and A2 several evenly spaced Cd$_{1-x}$Mn$_x$Te thin
layers were inserted during the growth of the 10~nm wide QW applying
the digital alloy technique~\cite{Kneip2006}.
In those samples the spin splitting can be described using
Eq.~(\ref{Eq04}).

\begin{figure}[htb!]  %01
\includegraphics[width=0.95\linewidth]{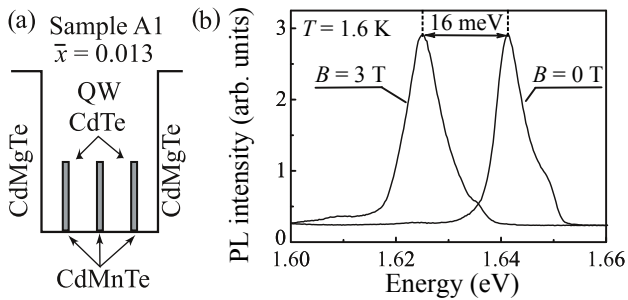}
\caption{ Design and PL data for sample\,A1 - (Cd,Mn)Te/(Cd,Mg)Te
DMS quantum well structure. (a) Sketch of the structure. (b)
Photoluminescence spectra at $B= 0$ and 3~T.} 
\label{Fig03}
\end{figure}

The sketch of  sample~A1 with three single monolayers of
Cd$_{0.86}$Mn$_{0.14}$Te is shown in Fig.~\ref{Fig03}(a).
Sample~A2 has similar design but is fabricated with two insertions
of three monolayers of Cd$_{0.8}$Mn$_{0.2}$Te.
In  the II-VI semiconductor compound the Mn atoms substitute the
Cd atoms and providing  a localized spin $S=5/2$.
In order to obtain a two-dimensional electron gas the structures
have been modulation doped by  Iodine donors introduced into the
top barrier at 15~nm distance from the QW. The electron density,
$n_e$, and mobility, $\mu$, obtained by magneto-transport
measurements,  as well as the effective average concentration of
Mn $\bar{x}$ and the Fermi energy $E_{\rm F}$, estimated from the
photoluminescence (PL) spectra, are summarized in
Table~\ref{table1}.
PL spectra obtained from sample~A1 at $B=0$ and 3~T
are shown in Fig.~\ref{Fig03}(b). Here the line for $B=3$~T is
substantially red-shifted (about 16~meV at $T = 2$~K) relative to that for zero field. 
This shift corresponds to 32~meV  giant Zeeman splitting of 
band states from which 6.4~meV fall into conduction band~\cite{Fur88}.

\begin{table}
\begin{tabular}{cccccc}
\hline
sample & \,\, $x$ & \,\,\,\,  $\bar{x}$ &\,\,\, $\mu$, cm$^2$/Vs &\,\,\,  $n_e$, cm$^{-2}$  & \,\, $E_F$, meV\\
\hline

A0  &  \,\,  0      &  \,\,  0      &   59000  &  4.2 $\times 10^{11}$  & 10.4 \\ %030107A - Sample C (A, CdTe)

A1  &  \,\,  0.14   &  \,\,  0.013  &   16000  &  6.2 $\times 10^{11}$  & 15.4 \\ %091906A - Sample B  (7-1)

A2  &  \,\,   0.20  &  \,\,  0.015  &   9500   &  4.7 $\times 10^{11}$  & 11.7 \\ %010408A - Sample A (C, 8-3)

\hline
\end{tabular}
\caption{Parameters of A0 - A2 samples. The effective average
concentration of Mn $\bar{x}$ is  estimated from the giant Zeeman
shift of the interband emission line. Mobility $\mu$ and electron
sheet density $n_e$ data are obtained at 4.2 K in the dark.}
\label{table1}
\end{table}

\begin{figure}
\includegraphics[width=0.99\linewidth]{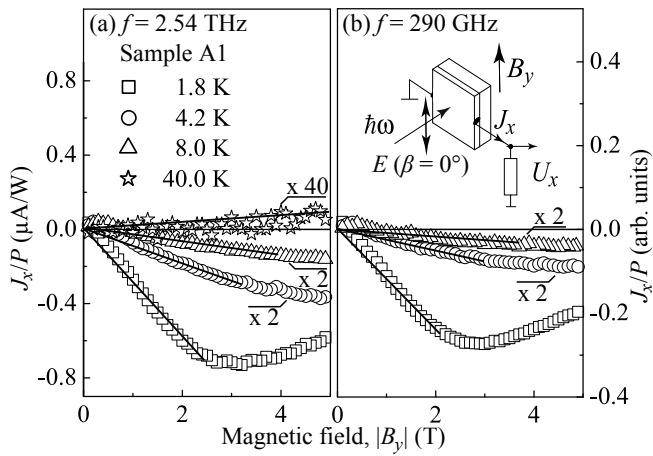}
\caption{Magnetic field dependence of the photocurrent measured in
(Cd,Mn)Te/(Cd,Mg)Te DMS QW sample A1 at  various temperatures and
applying (a) THz radiation,  $f = 2.54$~THz and (b) mw radiation,
$f=290$~GHz.
Solid lines are linear fits 
for low $B$. The inset shows the experimental geometry.}  \label{Fig04}
\end{figure}

We  start by describing the results  obtained with low power THz
and mw sources. The signal in unbiased samples is observed under
normal incidence with linearly polarized radiation for both
transverse and longitudinal geometries, where the current is
measured in the direction perpendicular, $J_x$, and parallel,
$J_y$, to the magnetic field $B_y$, respectively~\cite{footnote}.
Figure~\ref{Fig04} shows magnetic field dependence of the 
transverse photocurrent $J_x$.
The detected photocurrent is an odd function of the magnetic
field: It increases with raising magnetic field strength, vanishes
for $B=0$ and its sign depends on the magnetic field direction.
The signal linearly scales with the radiation power and does not
show a hysteretic behavior as ensured by sweeping magnetic field
from positive to negative fields and back (both not shown). For
convenience, in the discussion below we evaluate the data after
\begin{equation}
J_{x,y}(|B_y|)  = \frac{J_{x,y}(B_y > 0) - J_{x,y}(B_y < 0)}{2}\,\,\,,
\label{Eq08}
\end{equation}
which yields solely the strength of the magnetic field induced
photocurrent. Characteristic dependences of the photocurrent upon
variation of temperature, magnetic field strength, radiation
wavelength, intensity and polarization are the same for all
samples within each group and are qualitatively the same for all
DMS samples belonging to A1 and A2 groups. Thus, below we
consistently present the data obtained on one of the A1 sample.
Figures~\ref{Fig04}(a) and~\ref{Fig04}(b) show the  magnetic field
dependence of the transverse photocurrent $J_x /P$ measured in
sample~A1 under excitation with $cw$ THz  radiation ($f =
2.54$~THz) and mw radiation ($f= 290$~GHz), respectively. The
experiments reveal that at high temperatures, or at low
temperatures and moderate magnetic fields, the magnitude of
$J_{x}$ is  proportional to $B_{y}$, see~Ref.~[\onlinecite{footnote2}]. At low
temperatures and high magnetic fields, however, the photocurrent
saturates with increasing $B_{y}$. Moreover at $T = 1.8$~K a small
reduction of signal with increasing magnetic field is observed for
$B \gtrsim 3.5$~T.

\begin{figure}
\includegraphics[width=0.6\linewidth]{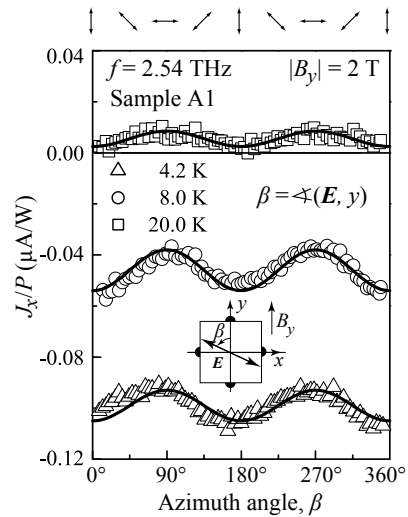}
\caption{Polarization dependences of photocurrent measured in
(Cd,Mn)Te/Cd,Mg)Te  DMS QW sample A1 at fixed magnetic field
$|B_y|= 2$~T and normal incidence of THz radiation ($f =
2.54$~THz) for $T=4.2$, 8, and 20~K. Fits are after
Eqs.~(\ref{Eq03}). The arrows on top show the orientation of the
light's electric field. The inset defines the azimuth angle
$\beta$.
 } \label{Fig05}
\end{figure}

Figure~\ref{Fig05} shows the polarization dependence of the
photocurrent measured in DMS sample~A1 excited by $cw$ THz
radiation. The data are obtained for $B_y = \pm 2$~T at which the
photocurrent does not show a saturation in the whole temperature
range, see Fig.~\ref{Fig04}(a). Consequently, the signal behavior
upon variation of the azimuth angle or temperature is not affected
by the photocurrent saturation. The current $J_x$ is well
described by the first equation of Eqs.~(\ref{Eq03}), see also inset in Fig.~\ref{Fig02}(b), and consists
of a polarization-independent, $ J_1$, and polarization dependent,
$J_2 \cos(2\beta) $,  components.
Following Eq.~(\ref{Eq03}) the individual contributions to the
transverse photocurrent, $J_1$ and $J_2$,
can be deduced from the experiment by taking, respectively, a
half-sum or  a half-difference of the signals obtained at $\beta =
0^\circ$ and $ 90^\circ$. In the longitudinal configuration we
detected only the polarization-dependent photocurrent $J_y =
J_3\sin(2\beta)$ well described by the second equation of
Eqs.~(\ref{Eq03}).

\begin{figure}
\includegraphics[width=0.7\linewidth]{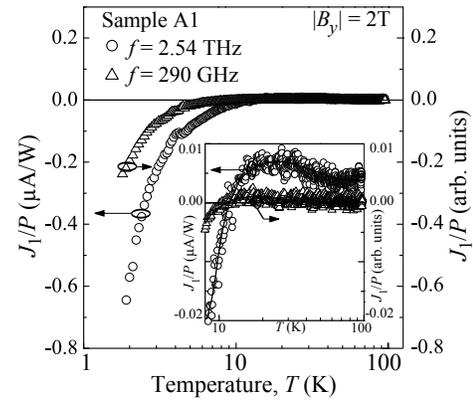}
\caption{DMS (Cd,Mn)Te/(Cd,Mg)Te sample~A1: Temperature dependence of
photocurrent  (polarization-independent) at  magnetic field $|B_y|
= 2$~T and  normal incidence of mw radiation ($f = 290$~GHz) and
THz radiation ($f = 2.54$~THz). The inset shows a zoom of $J_x(T)$
near the inversion point. }
\label{Fig06}
\end{figure}

The most striking observation comes from the investigation of the
temperature dependence of the polarization independent
photocurrent $J_1$. Figure~\ref{Fig06} reveals that a cooling of
the sample from 100~K down to 1.8~K results in, on the one hand, a
change of the current direction, and, on the other hand, an
increase of the photocurrent strength by about two orders of
magnitude. Such a temperature dependence is observed for both the
THz and mw radiation induced photocurrents  and the corresponding
data differs by a scalar factor only, see Fig.~\ref{Fig06}.
By contrast, in the reference non-magnetic sample A0, the drastic
enhancement of the signal magnitude and the inversion of the
photocurrent direction with the temperature decrease  have not
been observed (not shown).

\begin{figure}
\includegraphics[width=0.97\linewidth]{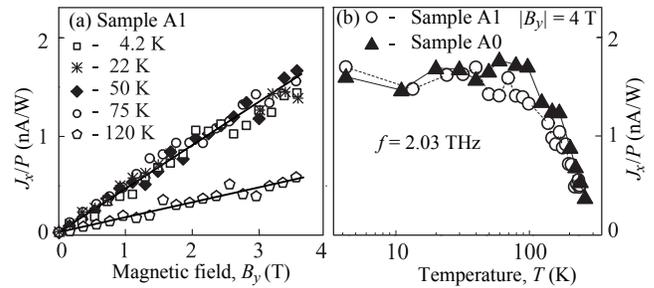}
\caption{(a) Magnetic field dependence of photocurrent at
different temperatures  for  DMS (Cd,Mn)Te/(Cd,Mg)Te sample~A1 and (b)
temperature dependence at fixed magnetic field for reference CdTe
sample~A0 and DMS (Cd,Mn)Te/(Cd,Mg)Te sample~A1 at normal incidence of
pulsed THz radiation ($f = 2.03$~THz) for with powers up to
$P\approx 40$~kW.  } \label{Fig07}
\end{figure}

The peculiar temperature behavior observed in DMS QWs excited by
low power radiation dwindle under application of high power pulsed
THz radiation with $P\approx 40$~kW, see Ref.~[\onlinecite{footnote3}]. While at low
power excitation the photocurrent direction changes upon cooling
and its magnitude strongly depends on the temperature
(Figs.~\ref{Fig04} and \ref{Fig06}) the current induced by high
power pulsed THz radiation neither undergoes an inversion nor
exhibit a significant dependence on $T$ in the range between 1.8
and 100~K, see Fig.~\ref{Fig07}. Furthermore, irradiation with
high power leads to a strong decrease of the magnitude of signal
normalized by the radiation power, $J_x/P$, compared to the one
for low power data ($\approx1~nA/W$ instead of $\approx1~\mu
A/W$). Moreover, the photocurrent saturation with increasing
magnetic field observed at low  power disappears, and the signal
excited by high power laser linearly scales with magnetic field
strength, Fig.~\ref{Fig07}(a). It is also remarkable that now DMS
samples and  non-magnetic samples show the same temperature
dependence: The photocurrent is almost independent of the sample
temperature below about 100~K and decreases
for $T>100$~K, see Fig.~\ref{Fig07}(b).

The experimental results described above are in a good agreement
with the theory of radiation-induced spin-polarized electric
currents in DMS quantum wells subjected to an in-plane external
magnetic field, see Sec. II. Comparison of the photocurrent
calculated after~Eqs.~(\ref{Eq01}) to (\ref{Eq04}) and shown in Fig.~\ref{Fig02} , 
with the
corresponding data, see Figs.~\ref{Fig04}~--~\ref{Fig06}, shows
qualitative similarity of the theoretical and experimental
results. In particular, the drastic enhancement of the
photocurrent magnitude and the change of its direction upon
samples cooling, as well as the observed saturation of the signal
with raising magnetic field strength are clear consequences of the
exchange interaction between the $s$-type conduction band
electrons and the half filled $d$-shell of the Mn$^{2+}$ ions.
The observed photocurrent sign
inversion upon temperature variation is caused by the  opposite
signs of the intrinsic and exchange Zeeman spin splittings, well
known for these materials. Due to the strong dependence of the
Brillouin function $\rm B _{5/2}$ [see last term in
Eq.~(\ref{Eq04})] on temperature sample heating
results in the rapid reduction of the exchange part to
the photocurrent and the dominance of the intrinsic one. The
interplay of intrinsic and exchange $g$-factors contributes also
to the deviation from the saturation behavior observed at low
temperatures. Here, instead of the saturation expected for the
Brillouin function a slight decrease of the photocurrent at high
magnetic fields is detected, see Fig.~\ref{Fig04}. Similar
behavior is seen for the calculated photocurrent shown by the
solid lines in Fig.~\ref{Fig02}(a), where both extrinsic and
intrinsic contributions are taken into account, see Ref.~[\onlinecite{footnote3a}].

While for low power  radiation the heating of the sample or the
manganese system  plays no essential role and the signal linearly
scales with radiation intensity a substantial increase of the
radiation power qualitatively changes the photocurrent formation.
Indeed, in the high power experiments neither an inversion nor a
photocurrent enhancement by cooling down the sample have been
observed, see Fig.~\ref{Fig07}. This indicates that at these
conditions the polarization of the Mn$^{2+}$ spins does not
contribute to the generation of current. Figure~\ref{Fig07}
demonstrates that the photocurrent in DMS samples excited by high
power radiation is at all temperatures proportional to the
magnetic field and varies with temperature in the same manner as
the one measured in non-magnetic  reference sample A0. It can be
well described with Eqs.~(\ref{Eq01}), (\ref{Eq02}), and
~(\ref{Eq04}) assuming vanishing contribution of the exchange
interaction. For low temperatures and degenerated electron gas the
characteristic electron energy $\bar{E}$ is equal to $E_{\rm F}$
and the photocurrent is nearly independent of~$T$. In the case of
a non-degenerated gas (higher temperatures) $\bar{E}$ is given by
$k_{\rm B} T$ and leads to a $1/T$ dependence of $J_x$. These two
regimes are clearly pronounced in Fig.~\ref{Fig07}(b) and, in
fact, are well known for spin-polarized photocurrents in
non-magnetic semiconductor structures~\cite{Nature06,Belkov2008}.

The observed photocurrent variation with the orientation of the 
radiation polarization
plane is also in agreement with the theory developed in Sec.~II.
The polarization dependence of the transverse photocurrent shown in Fig.~\ref{Fig05}
is in agreement with Eq.~(\ref{Eq03}) and corresponding calculated
curve shown in the inset in Fig.~\ref{Fig02}(b). It demonstrates
that this current is a result of superposition of the 
polarization-independent current due to energy relaxation of hot electrons
described by $j_1$
in Eqs.~(\ref{Eq03}) (relaxation contribution) and the
polarization-dependent one due to excitation given by the second
term in the first equation in Eqs.~(\ref{Eq03}). The %remained
longitudinal photocurrent is also observed and its polarization
dependence is in agreement with the second  equation in
Eqs.~(\ref{Eq03}).

The interplay of the giant exchange Zeeman splitting and the intrinsic one
in the total spin splitting  explains qualitatively the behavior
of the photocurrent upon changing magnetic field strength,
temperature, Mn doping as well as radiation intensity and
polarization. However, the observed increase of the current
strength at low temperatures  is substantially larger than the
giant Zeeman shift measured in the same structures by the
photoluminescence data. For example in sample~A1 at  $B=3$~T the
spin splitting, derived from PL data, changes from $-0.25$\,meV at
high temperatures (intrinsic value given by $g_e\mu_{\rm B}B$) to
2.6\,meV at 4.2 K and, hence, its magnitude swells by about a
factor 10. By contrast, the magnitude of the photocurrent at
$T=4.2$~K  increases by  about factor of~100  compared to that
measured for $T=40$~K, see Fig.~\ref{Fig04}(a). This quantitative
disagreement together with the strong temperature dependence of
the signal provide an evidence for the dominating contribution of
another DMS specific mechanism. This is the spin current due to
the spin-dependent electron scattering by polarized Mn$^{2+}$ ions
which was elaborated at the end of Sec.~II and is shown to amplify
the current conversion vastly.

\subsection{B. Mn-doped $p$-(In,Ga)As/InAlAs quantum well structures}

The second type of investigated samples III\,--\,V based DMS, with
Mn as the magnetic  impurity, are studied to a less extent than
the principal II\,--\,VI DMS family, but are already well
understood~\cite{DMS2010}. Currently this type of DMS structures are intensively
studied because of their prospect for spin-polarized carrier
injection~\cite{Ohno2010,Dietl2010,FlatteNature2011,Kohda06}  required  for spintronics applications. In III\,--\,V
semiconductors, like InAs or GaAs, Mn atoms substitute the group
III elements (In, Al or Ga), providing both  localized magnetic
moments with spin $S = 5/2$ and free holes~\cite{Dietl},
in contrast to II\,--\,VI materials, where Mn is an isoelectric impurity.

\begin{figure}[htb!]
\includegraphics[width=0.99\linewidth]{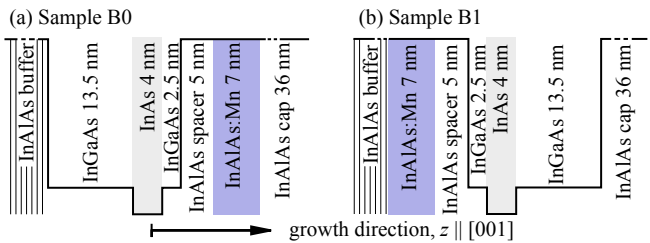}
\caption{Sketch of the Mn-doped (In,Ga)As/InAlAs samples with (a)
normal doped reference structure~B0 and (b) inverted doped DMS
structure~B1 for which segregation along growth direction results
in a Mn ion penetration into the InAs QW.  }
\label{Fig08}
\end{figure}

Samples investigated in the present work are compressively
strained InAs quantum wells  embedded in (In,Ga)As/InAlAs:Mn host
material with an In mole fraction of 75\% (for details, see Ref.~[\onlinecite{Wurstbauer09b}])
High mobility Mn modulation-doped single
QW structures were grown by molecular-beam epitaxy on
semi-insulating GaAs (001)-oriented substrates.
The layer sequences  of two fabricated
 Mn-doped samples are depicted schematically in
Fig.~\ref{Fig08}. The active layer consists of a 20~nm
In$_{0.75}$Ga$_{0.25}$As channel with an additional strained 4~nm
InAs QW, a 5~nm thick In$_{0.75}$Al$_{0.25}$As spacer, a 7~nm
thick Mn doped In$_{0.75}$Al$_{0.25}$As layer, and a 36 nm
In$_{0.75}$Al$_{0.25}$As cap layer. The samples differ in the
position of the Mn-doped layer. In the `normal' sample~B0, see
Fig.~\ref{Fig08}(a), a 5~nm In$_{0.75}$Al$_{0.25}$As spacer
followed by the Mn doping layer was grown after the InAs/InGaAs
channel, so that the InAs QW is free of Mn~\cite{Prechtl03,Wurstbauer09}.
In this sample the InAs QW is located 2.5~nm away from
the channel border, it is facing the Mn site and is separated from
Mn layer by 7.5~nm. The hole density and mobility obtained by
magneto-transport measurements are $n_h = 5.1 \times 10^{11}$
cm$^{-2}$  and $\mu = 8.6 \times 10^{3}$ cm$^2$/Vs. In the
`inverted' doped structures~B1, see Fig.~\ref{Fig08}(b), the
Mn-doped layer is also separated from the InAs QW by 7.5~nm but is
deposited before the channel growth. Due to segregation this
growth leads to a significant concentration of Mn ions in the InAs
QW.  The hole density in this
sample is $n_h = 4.4 \times 10^{11}$ cm$^{-2}$ and
 the mobility is reduced by at least a
factor of two compared to sample~B0.

\begin{figure}
\includegraphics[width=0.99\linewidth]{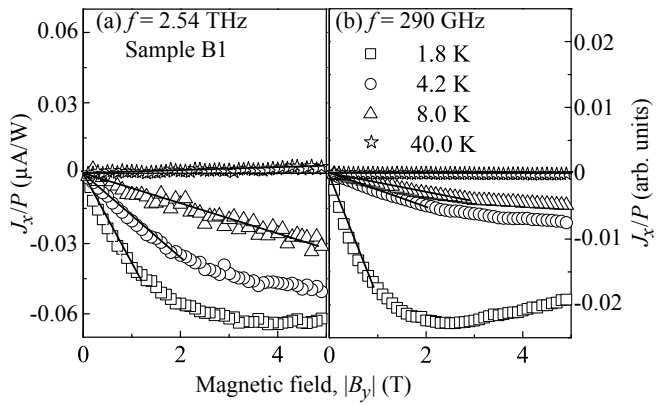}
\caption{Magnetic field dependence of photocurrent in inverted
Mn-doped (In,Ga)As/InAlAs DMS QW  at various temperatures. (a) and
(b) show photocurrent induced by THz radiation, $f = 2.54$~THz,
and mw radiation, $f=290$~GHz, respectively. Solid lines are 
linear fits for low $B$.} \label{Fig09}
\end{figure}

\begin{figure}
\includegraphics[width=0.8\linewidth]{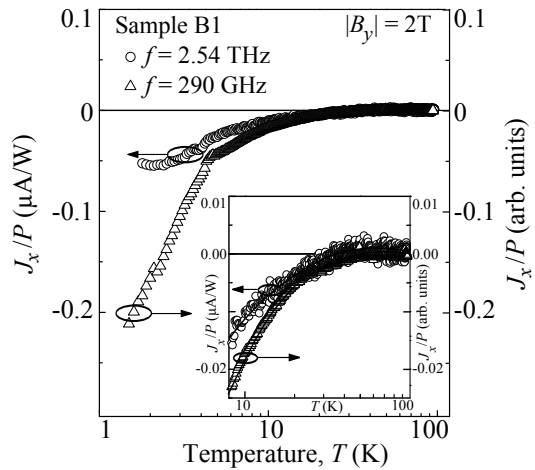}
\caption{Temperature dependence of photocurrent in $p$-doped
(In,Ga)As/InAlAs:Mn DMS QW  obtained at fixed magnetic field $|B_y|=
2$~T  applying mw, $f=290$~GHz, and THz, $f = 2.54$~THz,
radiation. The inset shows a zoom of the high temperature range.
} \label{Fig10}
\end{figure}

Figures~\ref{Fig09} and~\ref{Fig10} show the magnetic field and
temperature  dependences of the photocurrent generated in
sample~B1 under low power THz and mw  excitation. These data
reveal that the temperature decrease leads to a drastic
enhancement of the photocurrent magnitude as well as it changes
the linear in \textit{\textbf{B}} dependence of the signal into a
Brillouin-function-like saturation.~\cite{footnote4} These results
which are similar to that obtained in $n$-type (Cd,Mn)Te DMS
samples, are well described by Eqs.~(\ref{Eq01}), (\ref{Eq02}), and~(\ref{Eq04}), 
and provide a clear evidence for the exchange
interaction based origin of the observed photocurrent. The inset
in Fig.~\ref{Fig10} demonstrate that for mw-excitation raising
temperature does not result in the inversion of the current
direction. This result is expected for
$p$-type InAs DMS structures, in which, in contrast to $n$-type
II-VI QWs, the intrinsic $g_h$-factor for carriers
 and the exchange integral
have the same sign. For terahertz excitation a tiny positive
photocurrent is observed for $T \gtrsim 40$~K, which we attribute
to the interplay of the negative intrinsic spin photocurrent and
positive orbital photocurrent~\cite{Tarasenko08}.
Orbital photocurrent
may also be responsible for a weaker temperature
dependence of the THz radiation induced photocurrent compared to
the one excited by mw radiation. The photocurrent excited in the
normal Mn-doped sample B0, by contrast, is vanishingly small and
we do not observe any substantial increase of its magnitude upon
sample cooling. For both samples the signal is almost independent
of the orientation of the radiation plane. This observation
demonstrates that the photocurrent is dominated by the relaxation
mechanism.

Experiments applying high power pulsed THz laser radiation to both
B0 and B1 samples reveal that, similarly to the data obtained for
$n$-type (Cd,Mn)Te DMS samples [Fig.~\ref{Fig07}(a)], at all
temperatures the signal linearly increase with raising magnetic
field. Also the temperature dependence is very similar to that
detected in $n$-type (Cd,Mn)Te DMS samples [Fig.~\ref{Fig07}(b)].
The same results are obtained for the non-magnetic reference
\textit{n}-type InAs QW  sample doped by Si excited by high power
THz light as well by low power THz and mw radiation. All these
observations are in a good agreement with the theory of 
spin-polarized photocurrents in non-magnetic 
semiconductor structures, see Sec.~II.

While it was clearly observed in sample B1 at low power
excitation, at the first glance, the Mn doping outside of the
conducting channel  should not result in a magnetic behavior,
because the wave function of the carrier  does not penetrate to
the Mn location. However, the B1 sample is doped on the substrate
side and the Mn atoms penetrate towards the conducting channel due
to segregation of Mn atoms during the structure growth. The
segregation results in the presence of Mn$^{2+}$ ions in the
vicinity of the two-dimensional hole gas. The enhanced magnetic
properties manifest themselves by the colossal negative
magnetoresistance and the associated field-induced
insulator-to-metal transition observed in such structures~\cite{Wurstbauer10}.
By contrast, in the $p$-type InAs QW sample with Mn doping
on the surface side (sample~B0) the segregation shifts the Mn atom
distribution away from the 2D channel and the giant Zeeman
splitting of the hole subbands in InAs QWs is almost absent.
 The absence of the giant Zeeman splitting
in sample B1 substantiates the absence of residual Mn$^{2+}$ ions in
close vicinity to the two-dimensional hole gas. This further
verifies the interpretation of earlier magnetotransport
experiments~\cite{Wurstbauer09b}.

Similarly to II\,--\,VI DMS samples~A1 and A2 we observed that in
the inverted sample B1 the magnitude of the photocurrent measured
at 1.8~K is about two orders of magnitude larger than that at
40~K. Such enhancement is larger than that expected for the giant
Zeeman spin splitting and provides an indirect evidence for the
substantial contribution of the photocurrent due to scattering by
magnetic ions. However, the direct comparison of the current
variation to the Zeeman spin splitting is impossible, because no
PL or time-resolved Kerr rotation data for the InAs-based QWs are
in our disposal.

\subsection{C. Heterovalent \textit{n}-AlSb/InAs/ZnMnTe quantum wells}

InAs based DMS structures are usually characterized by \textit{p}-type conductivity~\cite{Wurstbauer09b,ohno1992}. 
Concerning \textit{n}-type In(Mn)As 
DMS only thin films and superlattices with 
mobilities in the order of 100 to 1000~cm$^2$/Vs have been reported so far~\cite{Munekata89,Soo1996,Acbas,Munekata03}.
The realization of  \textit{n}-type InAs based DMS QWs with high mobility and 
controllably exchange interaction remains an important issue.
A possible way to achieve this goal is to extend the heterovalent growth 
technology by the doping with magnetic ions.
While III-V and II-VI DMS systems are widely
studied and their magnetic properties are well
known, heterovalent \textit{n}-type AlSb/InAs/ZnMnTe quantum wells 
are new in the DMS family. 
These structures combine a narrow gap III\,--\,V QW with wide gap
II\,--\,VI barriers~\cite{Ivanov04}. Manganese is introduced into 
the ZnTe barrier
where it substitutes Zn and keeps electrically neutral providing
 a localized spin $S=5/2$. The enhanced magnetic properties
are caused by the penetration of electron wave function of
two-dimensional electrons into the (Zn,Mn)Te layer and can be
controllably varied by the position and concentration of Mn$^{2+}$
ions~\cite{Terentjev11}.

To fabricate  AlSb/InAs/(Zn,Mn)Te heterovalent structures with
Mn-containing barriers   two separate MBE chambers have been
applied, one for the III\,--\,V and the other for the II\,--\,VI
part. In the III\,--\,V MBE machine a buffer layer of GaSb
containing a strained AlSb/GaSb superlattice was grown. It follows
by a 4~nm thick AlSb barrier and a 15~nm thick InAs QW.

Before the first III\,--\,V part was transferred to the II\,--\,VI
MBE setup the structure was passivated ex situ by sulfur
exchanging a surface oxide, which then could be easily removed  to
start a coherent growth of ZnTe on top of InAs. In order to obtain
a diluted magnetic semiconductor barrier of InAs QW, sample C1, a
1~ML ($\approx$~0.32~nm)  MnTe  was introduced into the ZnTe
barrier at a 10 ML distance from the QW. 
By that, as a result of the segregation and 
diffusion processes, we obtain structure with Mn 
ions distributed over several monolayers of the surrounding 
ZnTe.  The maximum content of the remaining MnTe is estimated
to be well below 30 mol.$\%$.
Structure~C2 has the same
spacer with an adjacent 10~nm Zn$_{0.9}$Mn$_{0.1}$Te layer of
lower Mn concentration per ML. Sample~C0 is a reference
structure with non-magnetic ZnTe barrier.

\begin{figure}[htb!]  %03
\includegraphics[width=0.99\linewidth]{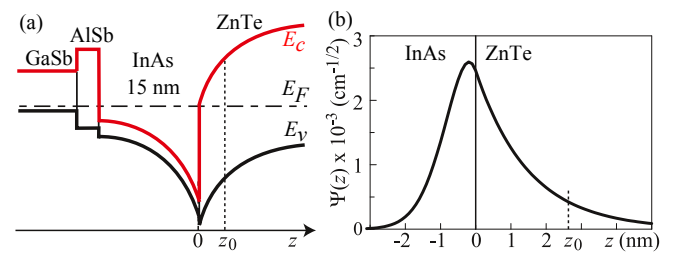}
\caption{ (a) Sketch of the band structure of hybrid AlSb/InAs/ZnTe 
samples, dotted line indicates the position of MnTe layer in  sample~C1.
(b) Electron wave function $\Psi(z)$ calculated for a triangular
QW with  the QW potential gradient $1.8 \times 10^7$~eV/cm
resulted from ionized donors at interface with the density $2
\times 10^{13}$~cm$^{-2}$, flat barriers, and the effective mass
$m^*=0.1m_0$. The latter corresponds to $m^*$ at conduction-band
bottom in ZnTe as well as in InAs with non-parabolicity being
taken into account. } \label{Fig11}
\end{figure}

The two-dimensional electron gas  has the density $n_e \sim  (1
\div 2) \times $10$^{13}$\,cm$^{-2}$ and the mobility $\mu \sim 5
\times $10$^3$\,cm$^2$/Vs at $T = 4.2$\,K. The most of 2D
electrons in hybrid QW originate from donor centers located at
III\,--\,V/II\,--\,VI heterovalent interface resulting in the
large surface density of positively charged donor centers at the
interface, while the Fermi level within the InAs layer is pinned
to that in the GaSb and ZnTe layers. Consequently, the structures
become highly asymmetric due to a strong built-in electric field.
The band structure of the sample C1 is sketched in
Fig.~\ref{Fig11}(a).

\begin{figure}
\includegraphics[width=0.99\linewidth]{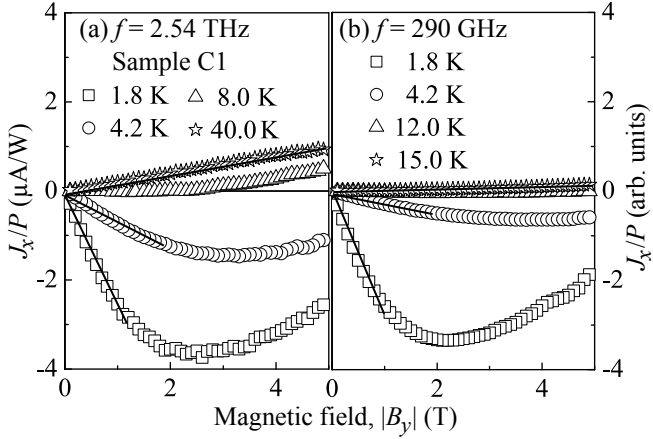}
\caption{ Magnetic field dependence of photocurrent measured in hybrid sample~C1 
at various temperatures and applying  (a)~THz radiation, $f =
2.54$~THz,  and (b)~mw radiation, $f= 290$~GHz.
Solid lines are fits for low $B$. } \label{Fig12}
\end{figure}

\begin{figure}
\includegraphics[width=0.99\linewidth]{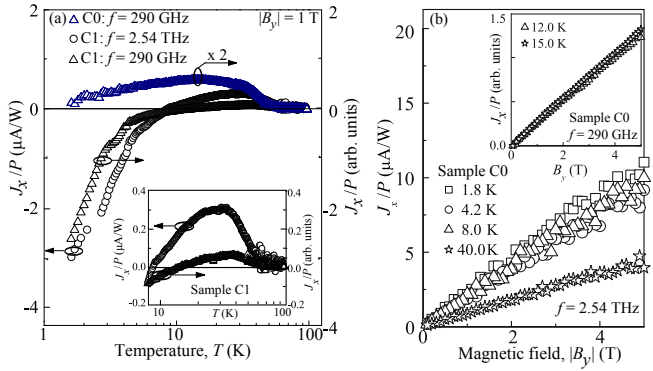}
\caption{(a) Temperature dependence of photocurrent measured in
samples C1 and C0  at fixed magnetic field $|B_y|= 1$~T applying
mw radiation, $f=290$~GHz, and THz radiation, $f = 2.54$~THz. The
inset shows a zoom of the high temperature region. (b) Magnetic
field dependence of photocurrent excited in hybrid non-magnetic
reference sample~C0  at various temperatures applying THz, $f =
2.54$~THz, and (in inset) mw radiation.
}
\label{Fig13}
\end{figure}

The magnetic field and temperature dependences of the photocurrent
induced in the DMS sample C1 are shown in Figs.~\ref{Fig12} 
and~\ref{Fig13}(a), respectively. Both plots demonstrate the
characteristic influence of Mn$^{2+}$ ions aligned by the external
magnetic field. The sign inversion of the photocurrent and the
strong enhancement of its magnitude by cooling the sample as well
as the non-linear magnetic field behavior (saturation at high $B$)
are clearly observed.~\cite{footnote5} The picture remains
qualitatively the same for both low power THz and mw radiations.
The only difference is the value of the inversion temperature
which is about 15~K for mw radiation induced photocurrent and
about 9~K for THz photocurrent. Figure~\ref{Fig13} also shows the
data for the reference non-magnetic AlSb/InAs/ZnTe QW sample~C0.
Here, in contrast to the sample~C1, the photocurrent shows linear
dependence on the external magnetic field in the whole temperature
range, it does not depend substantially on temperature for $T <
30$~K and for $T
> 30$~K decreases as $J \propto 1/T$. In sample~C2 with Zn$_{0.9}$Mn$_{0.1}$Te
 inserted in the barrier and distributed over a larger
distance from QW we observed less pronounced DMS properties (not shown). 
The photocurrent changes its sign upon cooling the samples at
$T\approx 2.5$~K, but at low temperature its magnitude is substantially 
lower than that detected in sample~C1. 

All these findings give a strong evidence for a substantial
influence of the exchange  coupling between the 2D electrons and the
Mn atoms introduced in the ZnTe barrier in sample~C1 and less
pronounced effect of magnetic impurities in the sample~C2. 
In both
magnetic samples~C1 and~C2 (Fig.~\ref{Fig11})  the 
Mn layers are separated from the QW by 10~ML thick spacer.
Therefore the exchange interaction is supposed to be mediated via a
penetration of the electron wave function $\Psi(z)$ into the
barrier.~\cite{Meilikhov_2010} The Zeeman splitting in structures
with Mn ions $\delta$-layer placed at $z=z_0$ can be estimated
using the standard expression
\begin{equation}
\label{Eq10} E_{\rm Z} = g_e \mu_{\rm B} B + \alpha_e N_{\rm{Mn}}
|\Psi(z_0)|^2 S_0 {\rm B}_{5/2}\left(\frac{5 \mu_{\rm B}
g_{\rm{Mn}} B}{2k_{\rm B} (T_{\rm{Mn}} + T_0)}\right) ,
\end{equation}
where $N_{\rm{Mn}}$ is the sheet Mn density. The necessary overlap
of the  electron envelope wave function with Mn$^{2+}$ ions is 
ensured by the strong asymmetry of the QWs due to the 
built-in electric field
discussed above. 
The calculations prove that the  wave function
deeply penetrates into ZnTe resulting in  the substantial 
overlap of $\Psi(z)$ and Mn$^{2+}$
ions in C1~structure, see Fig.~\ref{Fig11}(b). Due to the 
opposite signs of $g_e$ in InAs
and $\alpha_e N_{\rm{Mn}}$, under sample cooling the sign 
of $E_{\rm Z}$ inverses
resulting in the reversion of the photocurrent direction as
observed for C1 structure, see Figs.~\ref{Fig12} and \ref{Fig13}, 
and as well as for C2~samples. In the sample~C1
with Mn $\delta$-layer the current behavior at low temperature  is
dominated by the exchange interaction and almost follows the
Brillouin function: it is amplified by cooling the sample and, at
low temperatures, saturates with raising magnetic field [see
Figs.~\ref{Fig12} and~\ref{Fig13}(a)]. Estimations of the Zeeman
spin splitting in sample~C1 made after Eq.~(\ref{Eq10}) using
$N_{Mn}=10^{15}$ cm$^{-2}$ and $\alpha_e = 10^{-20}$~meV cm$^3$,
see~Ref.~[\onlinecite{Twardovski}], show that at $T = 1.8$~K and $B = 2$~T
exchange spin splitting should be one order of magnitude larger
than the intrinsic Zeeman splitting. This estimated value agrees
well with experimental findings, see Fig.~\ref{Fig13}(a), and
indicates that the Zeeman splitting based mechanism dominates the
current formation. 

The photocurrent data
obtained on the sample~C2 show much less pronounced magnetic
properties 
and give a further support of the suggested mechanism
for exchange interaction in C-type DMS structures. 
Indeed, because
of spatial distribution of the Mn over larger
distance from QW, in sample~C2
the overlap of the electron wave function with the Mn$^{2+}$ ions
should be substantially smaller than that in C1 structure. 

\section{V. Summary}

In summary, we demonstrate that the irradiation various types 
of low-dimensional diluted magnetic semiconductors by low power
terahertz or mw radiation causes spin-polarized electric current
if in-plane magnetic field is applied. Microscopically, the effect
originates from the spin-dependent asymmetric scattering of
carriers resulting in a pure spin current which is converted into
a spin-polarized electric current by magnetic field. Furthermore,
its behavior clearly reflects all characteristic features of the
exchange interaction and is giantly enhanced at low temperatures.
The spin-polarized electric current enhancement is caused by the
exchange interaction of carriers with Mn$^{2+}$ ions resulting 
in the giant Zeeman splitting. In the
structures with the Mn$^{2+}$ ions in the quantum well the 
efficiency of the current generation is
additionally amplified due to the spin-dependent scattering of
carriers by polarized Mn$^{2+}$ ions.  Our measurements carried
out on II\,--\,VI, III\,--\,V and hybrid II\,--\,V/II\,--\,VI  QW
structures doped by Mn demonstrate that the effect is very general
and can  be used for the efficient generation of spin-polarized
electric currents, e.g., applying conventional Gunn diodes, as
well as for the study of DMS materials. The latter could be of
particular importance for exploring DMS properties in materials
hardly accessible by optical or transport measurements.

\acknowledgments  The financial support from the DFG (SFB 689),
the Linkage Grant of IB of BMBF at DLR,  RFBR,   RF President grant
MD-2062.2012.2, and the Foundation ``Dynasty'' is gratefully
acknowledged. The research in Poland was partially supported by
the EU within European Regional Development Found, through grant
Innovate Economy (POIG.01.01.02-00-008/08).

\section{Appendix. Calculation of electron fluxes in the spin subbands}

Here we derive analytical equations for the electron fluxes
$\bm{i}_{\pm1/2}$ in the spin subbands.
We consider the excitation mechanism which is responsible for
polarization-dependent spin current. In the framework of kinetic
theory, the fluxes are given by
\begin{equation}\label{flux_eq}
\bm{i}_{s} = \sum_{\bm{k}} \bm{v}_{\bm{k}} \, f_{s\bm{k}} \:,
\end{equation}
where $s=\pm1/2$, $\bm{v}_{\bm{k}}=\hbar \bm{k}/m^*$ is the
electron velocity, $m^*$ is the effective mass, and $f_{s\bm{k}}$
is the electron distribution functions in the spin subband.

\subsection{A. Quasi-classical theory}

For low-frequency electromagnetic field, $\omega \ll
\bar{E}/\hbar$, the distribution functions of carriers in the spin
subbands can be found from the Boltzmann equation
\begin{equation}\label{kinetic_eq}
\frac{\partial f_{s\bm{k}}}{\partial t} + e \bm{E}(t)
\frac{\partial f_{s\bm{k}}}{\hbar \partial \bm{k}} = {\rm St}
f_{s\bm{k}} \:,
\end{equation}
where $\bm{E}(t) = \bm{E} \exp(-i \omega t) + \bm{E}^* \exp(+i
\omega t)$ is the electric field of radiation in QW, $\bm{E}$ is
the (complex) field amplitude, and ${\rm St} f_{s\bm{k}}$ is the
collision integral. In the case of spin-conserving elastic
scattering of electrons by impurities or structure defects, the
collision integral has the form
\begin{equation}\label{coll_int_eq}
{\rm St} f_{s\bm{k}} = \sum_{\bm{k}'} \left( W_{s\bm{k},s\bm{k}'}
\, f_{s\bm{k}'} - W_{s\bm{k}',s\bm{k}} \, f_{s\bm{k}} \right) \:,
\end{equation}
where $W_{s\bm{k},s\bm{k}'} = (2\pi/\hbar) \langle
|V_{s\bm{k},s\bm{k}'}|^2 \rangle \,
\delta(\varepsilon_{\bm{k}}-\varepsilon_{\bm{k}'})$ is the
scattering rate in the spin subband, $V_{s\bm{k},s\bm{k}'}$ is the
matrix element of scattering, $\varepsilon_{\bm{k}} = \hbar^2
\bm{k}^2 /(2m^*)$, and the angle brackets denote averaging over
scatterers. Taking into account $\bm{k}$-linear contributions to
the matrix element of scattering Eq.~(\ref{spin_scattering}), one
obtains, e.g., for the spin projections $\pm1/2$ onto the $y$
axis,
\begin{equation}\label{matrix_el_eq}
\langle |V_{s\bm{k},s\bm{k}'}|^2 \rangle = \langle V_0^2 \rangle +
4 s  \langle V_0 V_{yx} \rangle (k_x + k_x') \:.
\end{equation}
It is assumed in Eq.~(\ref{matrix_el_eq}) that the matrix elements
$V_0$ and $V_{yx}$ are real, $|V_{yx}| \ll |V_0|$, and $\langle
V_0 V_{yx} \rangle$ and $\langle V_0 V_{xy} \rangle$ are the only
non-zero components of the tensor $\langle V_0 V_{\alpha\beta}
\rangle$ in (001)-grown QWs. Below we suggest that the scattering
asymmetry is caused by short-range impurities or defects and,
therefore, $\langle V_0 V_{yx} \rangle$ is independent of the
directions of the wave vectors $\bm{k}$ and $\bm{k}'$.

To solve kinetic Eq.~(\ref{kinetic_eq}) we expand the distribution
functions in series of powers of the electric field,
\begin{equation}
f_{s\bm{k}} = f_{s}^{(0)}(\varepsilon) + [f_{s\bm{k}}^{(1)}
e^{-i\omega t} + {\rm c.c}] + f_{s\bm{k}}^{(2)} + \ldots \:,
\end{equation}
where $f_{s}^{(0)}(\varepsilon)$ is the equilibrium distribution
functions of electrons in the spin subband, $f_{s\bm{k}}^{(1)}
\propto |\bm{E}|$, and $f_{s\bm{k}}^{(2)} \propto |\bm{E}|^2$. The
first order corrections to the equilibrium distribution function
oscillate at the radiation field frequency $\omega$ and do not
contribute to dc fluxes. The directed fluxes $\bm{i}_{\pm 1/2}$ in
the spin subbands are determined by the second order in $\bm{E}$
corrections and obtained by multiplying $f_{s\bm{k}}^{(2)}$ by the
velocity and summing up the result over the momentum, see
Eq.~(\ref{flux_eq}). Such a procedure yields
\begin{eqnarray}\label{fluxes_final_eq}
i_{s,x} &=& M_{1,s} \, (|E_x|^2-|E_y|^2) + M_{2,s} \, |\bm{E}|^2 \:, \\
i_{s,y} &=& M_{1,s} (E_x E_y^* + E_y E_x^* ) + M_{3,s} i (\bm{E}
\times \bm{E}^*)_z \:, \nonumber
\end{eqnarray}
where
\begin{equation}\label{M1}
M_{1,s} = \frac{4 s e^2 \langle V_0 V_{yx} \rangle}{\hbar^4}
\sum_{\bm{k}} \frac{ \tau_{p} \, d (\tau_{p} \tau_{2} \,
\varepsilon^2)/d \varepsilon }{1+(\omega \tau_{p})^2} \frac{d
f_{s}^{(0)}}{d \varepsilon} \:,
\end{equation}
\vspace{-0.5cm}
\begin{equation}\label{M2}
M_{2,s} =  \frac{4s e^2 \langle V_{s} V_{yx} \rangle}{\hbar^4}
\sum_{\bm{k}} \frac{(1-\omega^2 \tau_{p} \tau_{2}) \,  \tau_{p}
\tau_{2} \, \varepsilon^2 \tau'_{p} }{[1+(\omega
\tau_{p})^2][1+(\omega \tau_{2})^2]} \frac{d f_{s}^{(0)}}{d
\varepsilon} \:,
\end{equation}
\vspace{-0.5cm}
\begin{equation}\label{M3}
M_{3,s} = - \frac{4se^2 \langle V_{s} V_{yx} \rangle}{\hbar^4}
\sum_{\bm{k}} \frac{ \omega \tau_{p} \tau_{2} (\tau_{p}+\tau_{2})
\, \varepsilon^2 \tau'_{p} }{[1+(\omega \tau_{p})^2][1+(\omega
\tau_{2})^2]} \frac{d f_{s}^{(0)}}{d \varepsilon} \:,
\end{equation}
where $\tau_{p}$ and $\tau_{2}$ are the relaxation times of the
first and second angular harmonics of the distribution function,
\[\tau_{p}^{-1} = (2\pi/\hbar)\sum_{\bm{k}'}\langle V_0^2 
\rangle (1-\cos\theta_{\bm{k}\bm{k}'})\delta(\varepsilon_{\bm{k}}-\varepsilon_{\bm{k}'}) \:,
\]
\[\tau_{2}^{-1} = (2\pi/\hbar)\sum_{\bm{k}'}\langle V_0^2 
\rangle (1-\cos 2\theta_{\bm{k}\bm{k}'})\delta(\varepsilon_{\bm{k}}-\varepsilon_{\bm{k}'}) \:,
\]
$\theta_{\bm{k}\bm{k}'}$ is the angle between $\bm{k}$ and
$\bm{k}'$, and $\tau_p'=d \tau_p / \varepsilon$. The fluxes
$\bm{i}_{+1/2}$ and $\bm{i}_{-1/2}$ are directed oppositely to
each other forming a pure spin current for equal distribution
functions in the spin subbands $f_{\pm1/2}^{(0)}$  and equal
scattering rates.

It follows from Eqs.~(\ref{fluxes_final_eq})-(\ref{M3}) that the
fluxes $\bm{i}_{\pm1/2}$ depend on the radiation polarization
state. For linearly polarized radiation, the dependence of the $x$
and $y$ components on the azimuth angle $\beta$ is given by
Eq.~(\ref{Eq03}) because $|E_x|^2-|E_y|^2 = - |\bm{E}|^2 \cos 2
\beta$ and $E_x E_y^* + E_y E_x^* = - |\bm{E}|^2 \sin 2 \beta$ for
the experimental geometry used. The term proportional to $M_{3,s}$
in Eq.~(\ref{fluxes_final_eq}) describes the contribution to spin
current that is sensitive to the sign of radiation helicity. It
can be excited by circularly or, in general, elliptically
polarized radiation and reversed by changing the sign of circular
polarization. Equation~(\ref{M3}) shows that the
helicity-sensitive current in QWs emerges due to the energy
dependence of the momentum relaxation time. If electrons are
scattered by short-range impurities or defects, then the energy
dependence of $\tau_p$ and $\tau_2$ can be neglected,
$\tau_p=\tau_2$, and the parameter $M_{1,s}$ takes the form
\begin{equation}\label{M_1n}
M_{1,s} = - 8 \frac{s\, n_{s}}{m^* \hbar} \frac{\tau_p^2 \,
e^2}{1+(\omega\tau_p)^2} \frac{\langle V_0 V_{yx} \rangle}{\langle
V_0^2 \rangle} \:,
\end{equation}
while $M_{2,s}$ and $M_{3,s}$ vanish. In this case, the excitation
mechanism leads only to the current contribution which is
sensitive to the linear polarization of radiation. The
polarization independent fluxes in the spin subbands are
completely determined by the energy relaxation mechanism.

\subsection{B. Quantum theory}

The presented above quasi-classical approach is not valid if the
photon energy $\hbar\omega$ is comparable or exceeds $\bar{E}$. In
this spectral range, an adequate microscopic theory of spin
current generation can be developed in the framework of quantum
consideration of intrasubband optical transitions in QW. Such
transitions are accompanied by electron scattering by impurities,
acoustic or optical phonons, etc., because of the need for energy
and momentum conservation. To first order in spin-orbit
interaction, the matrix element of optical transitions in the
subbands with the spin projection $s=\pm1/2$ along $y$ accompanied
by elastic electron scattering from short-range impurities has the
form~\cite{Nature06,Tarasenko06}
\begin{equation}\label{Matrix_optic}
M_{s\bm{k},s\bm{k}'} = \frac{e
\bm{A}\cdot(\bm{k}-\bm{k}')}{c\,\omega m^*}\, V_{s\bm{k},s\bm{k}'}
- 4s \frac{e A_x}{c\hbar} V_{yx} \:,
\end{equation}
where $\bm{A}=-i(c/\omega)\bm{E}$ is the vector potential of the
electromagnetic wave and $c$ is the speed of light. The first term
on the right-hand side of Eq.~(\ref{Matrix_optic}) describes
transitions $(e1,s,\bm{k}')\rightarrow(e1,s,\bm{k})$ with
intermediate states in the conduction subband $e1$, the second
term corresponds to the transitions via intermediate states in
other bands.

We assume that the radiation frequency is high enough, $\omega
\tau_p \gg 1$. Then, the distribution functions of electrons in
the spin subbands satisfy the kinetic equation
\begin{equation}\label{generation}
g_{s\bm{k}} = {\rm St} f_{s\bm{k}} \:,
\end{equation}
where $g_{s\bm{k}}$ is the optical generation rate,
\begin{equation}
g_{s\bm{k}} = \frac{2\pi}{\hbar} \sum_{\bm{k}',\pm}
|M_{s\bm{k},s\bm{k}'}|^2 (f_{s\bm{k}'}-f_{s\bm{k}})
(\varepsilon_{\bm{k}}-\varepsilon_{\bm{k}'} \pm \hbar\omega)
\end{equation}
and ${\rm St} f_{s\bm{k}}$ is the collision integral given by
Eq.~(\ref{coll_int_eq}). Taking into account spin-dependent terms
in the generation rate and the collision integral one can
calculate the distribution functions $f_{s\bm{k}}$ and then, using
Eq.~(\ref{flux_eq}), the fluxes $\bm{i}_s$. Such a calculation
yields
\begin{eqnarray}\label{i_quantum}
i_{s,x} &=& -  4 \frac{s \, n_{s} \kappa_{s} e^2}{\hbar m^* \omega^2} 
\frac{\langle V_0 V_{yx} \rangle}{\langle V_0^2 \rangle}  (|E_x|^2 - |E_y|^2) \:, \\
i_{s,y} &=& - 4 \frac{s\, n_{s} \kappa_{s} e^2}{\hbar m^*
\omega^2} \frac{\langle V_0 V_{yx} \rangle}{\langle V_0^2 \rangle}
(E_x E_y^* + E_y E_x^*)  \:, \nonumber
\end{eqnarray}
where $\kappa_s$ is a dimensionless parameter that depends on the
carrier distribution,
\[
\kappa_s = \frac{\int_0^{\infty} (1 + 2\varepsilon
/\hbar\omega)[f_{s}^{(0)}(\varepsilon) -
f_{s}^{(0)}(\varepsilon+\hbar\omega)] d
\varepsilon}{\int_0^{\infty} f_{s}^{(0)}(\varepsilon) d
\varepsilon} \:,
\]
and is equal to 1 and 2 for the cases $\hbar\omega \gg \bar{E}$
and $\hbar\omega \ll \bar{E}$, respectively. We note that, in the
frequency range $1/\tau_p \ll \omega \ll \bar{E}/\hbar$, both the
quasi-classical theory Eqs.~(\ref{fluxes_final_eq})
and~(\ref{M_1n}) and the quantum theory Eq.~(\ref{i_quantum}) give
the same result.


\begin{thebibliography}{99}
\bibitem{footnote0} Since the
average electron spin due to  the giant Zeeman effect is parallel
to $\bm{S}_{Mn}$, the directions of currents $\bm{j}_{\rm Z}$ and
$\bm{j}_{Sc}$ caused by the relaxation mechanism coincide for $u>0$.
\bibitem{footnote} We note that in the theoretical parts of the paper the current
density $j$ is used while in experiment the electric current $J$
is measured, which is proportional to the current density.
\bibitem{footnote2} At low magnetic fields $B_y \lesssim 0.5$~T a small deviation
from the linear coupling between $J$ and $B$ is detected, see
Fig.~\ref{Fig04}. The magnitude of this feature depends on the
contacts quality and,  most probably, originates   from the
illumination of contacts. However, this spurious signal is small
compared to that at high magnetic fields and will not be discussed
in the following.
\bibitem{footnote3} Note that the photocurrent pulses have typical duration  of about 100 ns, which corresponds
to the terahertz laser-pulses length.
\bibitem{footnote3a} An additional reason for this
discrepancy could be a contribution of the orbital
current~\protect \cite{Tarasenko08,Tarasenko11}, which, as discussed in Sec.~II, is almost insensitive to the enhanced magnetic
properties and is out of scope of this paper.
\bibitem{footnote4} A small deviation from the saturation behavior seen for microwave data we attribute to the
contribution of the orbital current discussed in Sec.~II.
\bibitem{footnote5}  Like in discussion of the results obtained for DMS samples of A and B type we attribute
a  deviation from the saturation behavior observed for high
magnetic fields to the contributions of intrinsic and orbital
photocurrents.
\end{thebibliography}
\end{document}